\documentclass[twocolumn,showpacs,preprintnumbers,amsmath,amssymb,showkeys]{revtex4}
\usepackage[pdftex]{graphicx}
\usepackage{amssymb}
\usepackage{bm}
\def\lesssim{\mathrel{\hbox{\rlap{\hbox{\lower4pt\hbox{$\sim$}}}\hbox{$<$}}}}
\def\gtrsim{\mathrel{\hbox{\rlap{\hbox{\lower4pt\hbox{$\sim$}}}\hbox{$>$}}}}

\begin{document}
\setcounter{page}{1}
\title[]{Active Galactic Nuclei: Unification, Blazar Variability and the Radio Galaxy/Cosmology Interface}
\author{Paul J. \surname{Wiita}}
\email{wiita@chara.gsu.edu}
\thanks{Fax: +1 404-651-1389}
\affiliation{Department of Physics and Astronomy, Georgia State University,
Atlanta, 30302-4106, USA }
\date[]{Received March 15 2006}

\begin{abstract}
We first review some basic properties of the most important classes of active galactic
nuclei (AGN), including quasars, blazars, Seyfert galaxies and radio galaxies.  The most
commonly accepted type of scheme designed to individually unify the radio-loud and
radio-quiet categories of AGN is based upon
three parameters: black hole mass, accretion rate, and our orientation to the accretion flow.
Some recent evidence from optical
microvariability of several classes of AGN points in favor of  a strong unification scheme that unites both radio-loud
and radio-quiet categories.   An important question concerning the
nature of blazars and other AGN whose jet emission appears to dominate their
spectral energy distributions involves the velocities of those flows.  A variety
of apparently contradictory observations can be reconciled if such flows are 
ultrarelativistic but have an opening angle of a few degrees.  Radio galaxies (RGs) were
much more numerous at redshifts $\sim$2 than they are today.  Combining this
fact with the realization that older RGs at such redshifts are very difficult to detect,
and with cosmological simulations of the growth of structure in the universe has led us to propose that
RG lobes have impacted a significant fraction of the cosmic web of baryons.
These impacts may have triggered extensive star formation and perhaps even engendered
new galaxies; they also
probably played important roles in the spreading of magnetic fields and 
heavier elements into the intergalactic medium.
\end{abstract}



\maketitle

\section{INTRODUCTION}

As astronomers have probed the centers of galaxies more deeply, with ever larger
telescopes, and more broadly, with measurements ranging from the radio through the gamma-ray
bands, the mysteries hidden there are slowly being revealed.  It is now clear that
a substantial fraction of, and very probably essentially all, galaxies house a
supermassive black hole (SMBH) in their cores.  As there seems to be a rough proportionality
between the mass of these SMBHs and of the stars which comprise the spheroidal component
of those galaxies, e.g., \cite{geb00}, the idea that there is some intimate feedback between the
formation of galaxies and their central BHs has become widely accepted over the past few years,
although the exact nature of this feedback is currently being actively debated.
At any given time, a small fraction of these SMBHs behave as powerhouses, anchoring
accretion disks and jets which can produce the spectacular emissions arising from
active galactic nuclei (AGN).

In this paper I shall only note a few aspects of the manifold phenomena associated with AGN
and shall focus on some recent research which helps to tie together different classes of AGN.
Section II contains a (very) brief tour of the zoo of beasts which together are usually classified
as AGN.  In \S III a summary of the most common scenario which has been proposed to
unify these apparently different animals is presented, though it must be stressed that this 
picture cannot be a complete one and there
are several viable variations on this theme.  There are many excellent reviews of 
AGN, including three fairly recent monographs \cite{pet97} \cite{kro99} \cite{kem99}. Please note that for reasons of space
the list of references given in this rather broad ranging paper is quite sparse: nearly every one should
be taken to include the prefix, ``e.g.'', and the suffix, ``and references therein''.
Section IV describes some of our work on
intranight optical variability (or microvariability) which supports the idea that very similar processes
are responsible for these fluctuations in several different types of AGN, and thereby strengthens
the unification scheme.  Section V shows how some apparently discrepant observations of
certain blazars can be understood if a simple modification to the usual jet picture is introduced
and \S VI discusses how radio galaxies can play a substantial role in triggering extensive star
formation.  A few key conclusions are given in \S VII.

\section{TYPES OF ACTIVE GALACTIC NUCLEI}

We can usefully begin our discussion of AGN by separating galaxies with truly active nuclei from pretenders to that title.
Many galaxies contain large zones of star formation emitting a great deal of energy, but the activity in these 
starburst galaxies usually arises over length scales of several hundred parsecs.  Other galaxies are
disturbed by recent interactions, and their tidal tails or ring-like zones of enhanced star formation
are spectacular, but they too do not meet our criterion of substantial activity concentrated in the very
central region of the galaxy.  In a true AGN much of the activity can be localized to a region $<$ 1 pc in size, and in the rest of this section we give the salient properties of the main classes of AGN beginning with the first type to be recognized.

{\bf Seyfert galaxies} are distinguished morphologically by a bright semi-stellar nucleus embedded in a
spiral galaxy (usually of Hubble morphological type Sa or Sb) which emits roughly
as much energy in visible wavelengths as do all of the stars in that galaxy, i.e., $\sim 10^{44}$ erg s$^{-1}$.
This class of AGN was first recognized by Carl Seyfert \cite{sey43} and the spectra he took revealed 
strong nuclear emission lines, many of which were often very broad (up to 8500 km s$^{-1}$, full width at zero intensity).
Later examination of these Seyfert (Sy) galaxies \cite{kha74} showed that they could be divided into two basic classes:
the Type 1 Sy galaxies, where these broad lines were present, but narrower lines with widths of several hundred km s$^{-1}$
were also seen for the same transitions as well as others, and the Type 2 Seyferts, which  
only showed the narrower lines.  
These narrow lines were
still significantly broader than those found in normal galaxies and detailed analyses showed that they were emitted by
partially ionized gas clouds of low densities, $n_e \approx 10^{3-6}$ cm$^{-3}$.  The broad lines were only seen for
permitted lines, and the lack of the corresponding forbidden lines implied that the gas clouds emitting them were of substantially
higher densities, $n_e \gtrsim 10^9$ cm$^{-3}$.   The higher velocities of the broad line region (BLR) were taken to imply that
the clouds producing that emission were moving faster, and were presumably in a stronger gravitational potential.

Other key characteristics of all Seyfert galaxies include:  (a) substantial variability in the optical band;  (b) significant and often rapidly variable x-ray emission, though Sy 1's are much stronger hard x-ray emitters than are Sy 2's; (c) usually very little radio emission, though sensitive measurements can usually find a weak radio source coincident with the optical nucleus.  The quasi-thermal starlight in the central regions is usually overwhelmed by a ``featureless continuum'' in the Sy 1's, whereas this continuum is more
difficult to detect in Sy 2's.  The fact that the x-ray emission can fluctuate by factors of two in the course of hours in many Seyfert galaxies implies that the length scales of these emitting regions are only light-hours.  

The sizes of the BLRs have been determined for several Sy 1's through the technique of
``reverberation mapping'' \cite{pet97}.  Here the time lags between changes in the continuum flux and those in several different lines, corresponding to different ionization potentials, are carefully measured.  Typical lags are a few days for lines such as
N$~$V $\lambda$1240, several days for He II $\lambda$4686 and over a dozen days for H$\beta ~\lambda$4861, implying that the
more strongly ionized gas is closer to the continuum source and the distances involved are $\sim 0.01 -0.1$ pc.  All indications are that the narrow line regions (NLR) lie at much greater distances, $\gtrsim$ 10 pc.  A few per cent of spiral galaxies are Seyferts,
implying that either all galaxies go through this phase for a few percent of their lifetimes
or that only a small fraction of galaxies spend a larger part of their lives in such an
active state.

The next class of AGN to be studied were {\bf radio galaxies} (RGs), which were first noticed as pairs of radio emitting lobes straddling
optical galaxies in the 1950s.  As the radio maps improved it became clear that these extended lobes of the Fanaroff-Riley Type II (FR II, \cite{fan74}) RGs
usually contained hot-spots which were being energized by flows that could occasionally be seen as jets.  Radio galaxies with most of their emission coming from
the jets themselves and concentrated toward the middle of the source were usually weaker and are usually called FR I's.  The great majority of all extragalactic radio sources are associated with elliptical galaxies, and while some of those elliptical hosts are peculiar, many are not optically exceptional.  Although the sizes of these RGs vary immensely, from total linear sizes of less than 1 kpc to over 1 Mpc, sizes of $\sim$300 kpc are typical.  With the advent of Very Long Baseline Interferometry (VLBI) these jets could often be traced into the cores of the galaxies, 
and the ``knots'' in these jets often appeared to be moving outward from the very center \cite{bbr84}.  

The quasi-power-law spectra and high polarizations of the radio emission imply that we are seeing synchrotron radiation emitted by extremely relativistic particles (electrons and perhaps positrons too) moving in partially ordered magnetic fields of $\sim 10^{-4}$ gauss in the hotspots
and $\sim 10^{-6}$ gauss in the lobes.  The very large volumes of these regions implied that total energies of $\gtrsim 10^{60}$ erg were not uncommon.  The large sizes demanded that these RGs had to be turned on for many millions of years, though determining precise velocities of expansion and ages are impossible.  All of these extreme properties mark RGs as the largest and most energetic connected objects in the universe.

The VLBI components of RGs sometimes appear to be moving at superluminal velocities, which can be best understood as a special-relativistic effect which arises if the jet is pointing within a few degrees to our line-of-sight to
the active core (\S V.A).  The launching of new VLBI knots usually corresponds to an increase in radio flux, and  the
temporal evolution of radio emission over a range of frequencies can be nicely understood in terms of shocks propagating down
relativistic jets \cite{aah90}.

The discovery of {\bf quasars} in the 1960's brought the study of AGN to a level of prominence in astronomy which it has retained ever since, e.g., \cite{wii85}.  While the first quasars were discovered by finding 
apparently stellar optical counterparts to strong radio sources (hence, the catchy quasi-acronym,
``QUAsi-StellAr Radio Source'', it was soon realized that many other highly luminous and apparently compact optical sources did not have significant radio counterparts, and this led to the introduction of the more general term, quasi-stellar-object, or QSO, with QSR explicitly denoting a ``quasi-stellar radio'' source.   When the high redshifts of QSOs were understood their immense powers were deduced: optical luminosities in excess of $10^{46}$ erg s$^{-1}$ are normal for a QSO.  It is currently believed that $< 10\%$ of all QSOs are radio-loud (i.e., quasars in the formal sense) \cite{ive02}, but all of them have very substantial emission in the infrared, ultraviolet and x-ray bands to go along with their huge optical outputs.

The key feature of QSO spectra was a great similarity to that of the Type 1 Seyfert galaxies:
extremely broad emission lines.  Very careful studies eventually revealed ``fuzz'' around the QSOs which corresponded to galactic emission with the same redshift as that of those QSO lines, and so
the consensus arose that QSOs, at least the radio-quiet QSOs (RQQSOs), were fundamentally
Seyferts on steroids.  Instead of having an extremely bright Sy nucleus with a power comparable to that of all the stars in the galaxy, a QSO has  phenomenally bright nuclear emission which substantially outshines all of those stars, making the galaxy difficult to even detect.  The spectral energy distribution (SED) of most QSOs could roughly be described as a power-law: $F_{\nu} \propto \nu^{-1}$ between the IR and UV bands, though there was often evidence of a ``big blue bump'', or quasi-thermal component, in the (rest-frame) near-UV.  A subset of QSOs also exhibit very broad absorption line troughs (corresponding to velocities up to 30,000 km s$^{-1}$), implying the presence of strong outflows which are optically thick in these BALQSOs.
  
As QSOs continued to be studied, they were also seen to be variable, though usually not as dramatically (in a fractional sense in a given length of time) as are the Seyfert galaxies.  Nonetheless, changes of $10\% $
or more over the course of a year were commonly seen, and a subset of Optically Violently Variable (OVV)
quasars were identified; these QSOs might change their output by a factor of two over a year or by 0.1 magnitudes over a few nights.  While the optical polarization of most QSOs is very small, a subset showed optical linear polarization exceeding $3\%$ or so, indicating a significant contribution from
synchrotron emission.  This High Polarization Quasar (HPQ) subset is nearly isomorphic to the OVV
group.

{\bf BL Lacertae} (BL Lac) objects are the fourth and last discovered of the main classes of observationally specified AGN.  These were first thought to be very variable stars (hence the nomenclature) lacking usual stellar absorption lines,
but in the late 1970s they were recognized to be powerful radio and x-ray sources, and hence, quite clearly non-stellar.  However, their spectra are essentially featureless,
so that redshifts, thence distances and luminosities, were, and are still, very difficult to estimate.
But, as with the discovery of the galaxies in which quasars are embedded, more careful study allowed for the discovery of a few weak emission lines in some BL Lacs, and thus
cosmological redshifts and high apparent powers were attributed to them.

The key property of all BL Lacs is that the strong variability extends throughout the
electromagnetic spectrum, and the advent of gamma-ray astronomy showed that many BL Lacs
were also powerful (but variable) high energy photon sources.  This very rapid variability implied a very compact origin of most of the emission.  One other important property of the BL Lacs is their high polarization in the optical, as well as radio, band.  

There appeared to be significant differences in the SEDs and redshift distributions of BL Lacs discovered in radio and x-ray surveys, so there was much discussion of RBLs vs.\ XBLs in the literature in the 1980s and early 1990s, but it now is clear that there are also many BL Lacs with intermediate types of SEDs.  The key point is that all BL Lacs seem to have SEDs with two humps, with the one at lower frequencies dominated by synchrotron emission and the one at higher frequencies arising from the
inverse Compton process.  The locations of the peaks of those maxima can be substantially different, depending on the maximum energies of the relativistic particles, and this gives rise to the apparent RBL/XBL dichotomy \cite{xxx01}.  

Because BL Lac objects, OVVs and HPQs all demonstrate rapid variability and high polarization, these types of AGN are now usually grouped under the common rubric of
{\bf blazar}.

\section{UNIFICATION SCHEMES}

For all types of AGN, the emission was seen to arise from the  center of the host galaxy,
and in the case of those with compact radio cores, VLBI could pin this down to roughly parsec scales that seemed to coincide with the very center of the galaxy.
The variability seen in all types of AGN strongly supported the notion that their
energies were mainly released on small scales, producing light travel times across them
corresponding to those observed timescales of minutes to years.  The immense energies
involved demanded very efficient powerhouses, since the brightest quasars implied the
complete conversion into energy of many solar masses per year.  Together, these points
(and others) strongly implied that all AGN are somehow powered by accretion onto supermassive black holes, since: BHs are the most compact regions imaginable; BHs would be expected to exist at the centers of galaxies;  accretion onto BHs can be very efficient, converting between 0.056 and 0.32 of the infalling mass into radiation and mechanical outflows (for standard thin accretion disks, depending upon whether the BH is non-rotating or maximally rotating) \cite{tho74}.

For example, if a SMBH has a mass of $10^8$M$_\odot$, its Schwarzschild radius, $R_s = 2GM_{BH}/c^2 = 3 \times 10^8$km $\simeq 1 \times 10^{-5}$pc.  Temporal variations down to a scale of $\Delta t =  R_s/c \simeq 10^3$s would then be reasonable.  If an accretion rate, ${\dot M} = 10$M$_\odot$/yr of matter falls onto such a SMBH at an efficiency, $\epsilon = 0.06$, then $L \simeq 3 \times 10^{47}$ erg s$^{-1}$, a typical quasar power, can be generated.  Using the approximate relation for energy released falling into a gravitational pit, 
$L \sim G M {\dot M}/R$, substituting in the above values of $M_{BH} and {\dot M}$, and equating $R$ to $R_s$ (since most of the radiation is emitted within a few times the
Schwarzschild radius) one gets agreement to better than a factor of two, thereby illustrating the strong plausibility of this scenario on spatial, temporal and energetic grounds \cite{wii85}..

Currently, the weakest form of unification, the hypothesis that all AGN are powered by SMBHs, is completely accepted.  Going beyond this basic point things become more controversial, but the ideas that all (or at least most) radio-loud AGN can be unified through an orientation based scheme where jet direction plays a key role \cite{bla78},\cite{bar89}, and that all (or at least most) radio-quiet AGN can be unified through an orientation based scheme where an obscuring torus (or something similar) plays a key role \cite{ant83}, are also widely accepted.

\subsection{Radio Quiet AGN}
  
The clinching evidence that the differences between Sy 1 and Sy 2 galaxies could mainly be explained by different viewing angles came from the discovery that the polarized emission of some Sy 2 galaxies did show the presence of broad lines \cite{ant80} buried under the much stronger narrow lines.  This was interpreted as meaning that even though one couldn't directly see the BLR in an Sy 2, it was still present, but usually hidden by some optically thick torus on a scale of $\sim$1 pc, i.e., bigger than the BLR but smaller than the NLR, so that both Sy 2's and Sy 1's
evince narrow lines, but only Sy 1's allowed us to see their BLRs. But if there was enough of an ionized electron cloud above the axis of this torus then it might scatter (and polarize) a small fraction of the BLR and continuum emission into our line of sight.  An obscuring torus (which is probably comprised of many smaller cold clouds containing dust within a roughly toroidal zone) would also explain why the non-thermal optical continuum was not present or much weaker in the Sy2's and why they were much weaker x-ray sources as well \cite{ant83}.  Statistical studies indicate that there are between 2 and 3 times as many Sy 2's as Sy 1's.  This implies that only if we are looking within 
$\sim40^{\circ}$ of the axis of the torus (which is presumably along the spin axis of the BH) will we see such a modestly powered AGN as a Sy 1, for larger viewing angles intercept the torus and allow us to only see a Sy 2.

As noted above, the main difference between RQQSOs and Seyfert 1s is the greater overall power of the QSO, since their spectra are very similar.  A secondary difference is the longer timescale over which detectable
luminosity changes are observed.  Thus we can unify these two classes of radio-quiet AGN merely by saying that the QSO involves higher accretion rates onto more massive BHs, and that for both Sy 1's and QSOs we are able to see the continuum source (with a direct contribution from the accretion disk probably producing the big blue bump) and the BLR.  All masses and dimensions are bigger for QSOs than for Sy's (say, M$_{BH,Sy} \sim 10^{6-7}$M$_\odot$, while
M$_{BH,QSO} \sim 10^{8-9}$M$_\odot$) in this picture.  

The long lasting problem for this
radio-quiet unification scenario has been the lack of obvious equivalents of Type 2 Seyferts among the most luminous objects.
If it is to hold, there should be roughly twice as many ``Type 2'' quasars with only narrow lines visible as there are quasars, yet for quite some time there were no such creatures known.  The discovery of ultra-luminous infrared galaxies
(ULIRGs) provides a viable candidate for these Type 2 quasars, since the immense IR powers
($\sim 10^{12}L_{\odot}$) found for these objects could well be (at least partially)
due to the optical/UV emission from a QSO being absorbed by a dusty torus which then reradiates it in the IR \cite{xxx02}.  However, there is substantial evidence that most of the emission of many ULIRGs is dominated by more extended star forming regions, and that
any AGN contribution often may be small.  Still, since the ULIRG population appears to be high, even a fraction of them may comprise the Type 2 QSOs.  Deep x-ray and sub-mm surveys also appear to be finding optically faint sources which may be the missing Type 2 quasars, e.g.\ \cite{xxx03}.  

An alternative (or supplement) to the orientation based unification scheme is a unification scheme more determined by temporal evolution.  It is certainly possible that the same active nucleus will have a radically different appearance over long periods of time as it accretes at different rates and has different densities of gas in its vicinity.  Recent work on
growth of BHs which incorporates strong feedback from the BH powerhouse on the galactic
gas can nicely explain many things, including the evolutions of the QSO luminosity function over cosmological times \cite{hop05}.

\subsection{Radio Loud AGN}

The observation that the mean projected
linear size of quasars was smaller than that of FR II RGs of similar high radio luminosities led to the suggestion that
they were identical beasts, with the former being viewed within an angle of $\sim 45^{\circ}$ to the jet direction, while the FR IIs were being seen closer to the plane of the sky \cite{bar89}.  The larger number of RGs than QSRs in low frequency selected radio catalogs (where most of the emission comes from the extended lobes) is in accord with this idea for such a demarcation angle \cite{urr95}. Further, the more frequent detection of strong radio cores in quasars than in RGs is consistent with this picture, since Doppler boosting of a jet pointing relatively close to us would magnify the emission from the center in the case of quasars (see \S\S IV and V for some details).  The apparent superluminal motions seen in many quasars (\S V) but not in FR II RGs also fits in with this jet orientation based unification hypothesis, so it is now widely accepted.  Radio-loud quasars seen at the smallest angles to the jet axis then should correspond to most core-dominated quasars (CDQs) since the Doppler boosted central emission could then be substantially brighter than the intrinsically more luminous, but quasi-isotropically emitting, radio lobes. At intermediate powers, many of the FR II broad line RGs are taken to be the objects viewed at modest angles to the line-of-sight while the narrow line RGs
are those seen at greater angles \cite{lai94}.

As for the weaker radio sources, the BL Lacs are now commonly taken to be the Doppler boosted counterparts of the FR I RGs.  The statistics of these populations as seen at different radio frequencies are nicely explained if the BL Lacs are really FR I's viewed within a narrow angle of $\sim$10$^{\circ}$ to the line-of-sight \cite{urr95}.  If this is the case, and it almost certainly is, the BL Lacs are actually intrinsically weaker than other AGN classes, but their observed emission is dramatically boosted by special relativistic effects.
Some of the more powerful blazars, such as the OVVs, are probably better considered to be produced by FR II RGs and are very thus very well aligned quasars.  Differences in the
observed optical lines (narrow line RGs vs broad line RGs)  are in rough accord with the idea that
the obscuring tori also exist in radio-loud AGN, but this is one of the least well established aspects of the unification scenario.  Another poorly understood aspect of the
nature of radio-loud AGN is the reason for the differences between FR I and FR II classes,
though the power of the jet compared with the nature of the ambient medium through which it propagates probably plays the major role \cite{gkw00}.  An even more important question is: why are some AGN radio-loud whereas most are radio-quiet?  While there have been many proposals put forward to explain this fundamental dichotomy, quite a few involving the nature of the accretion flow, none is really satisfactory, e.g., \cite{har06}.

It should be stressed that the basic unification schemes in which there is a single
critical angle dividing Sy 1s from Sy 2s
and another pair of critical angles dividing QSRs from FR IIs and BL Lacs from FR Is
is very probably correct to zeroth order, but is also definitely an oversimplification.
The statistics of QSR/FR II jet lengths at different redshifts can only be explained if
more powerful RGs have larger critical angles \cite{gkw96} and the x-ray properties
of Type 1 and Type 2 radio quiet AGN seem to require a wider torus opening at higher luminosities \cite{tri06}.  Differences in their x-ray properties also seem to imply that the so-called ``low-excitation RGs'' cannot be simply unified with the other classes of RG \cite{har06}.

\section{MICROVARIABILITY OF DIFFERENT AGN CLASSES}

One important way of understanding the relation between radio quiet and radio loud AGN involves examining variations in their optical emission.  The advent of CCD cameras allowed accurate
differential photometry on sub-hour timescales, and the reality of such small (a few percent, or a few hundredths of a magnitude) microvariability for blazars was then convincingly established \cite{mil89}.  
These fast temporal variations provide unique probes of AGN, as they correspond to the smallest accessible physical scales of these
central engines.  

Monitoring of a variety of different types of AGN by groups based on every continent showed that radio-loud quasars also evinced
microvariability, although less frequently than did blazars, e.g., \cite{dut96}.   The relative proximity of Seyfert galaxies makes them brighter than most quasars, but the rapidity of changes in atmospheric seeing adds varying amounts of galactic light to their central regions, making the ground based optical microvariability results obtained for Sy's questionable.  Since quasar and blazar emission is completely
dominated by the active nucleus, differential comparisons of the QSO light-curve with those of at least two nearby stars of similar magnitudes and colors enables the extraction of
real variations of $\lesssim 0.01$ mag, even in the presence of variable seeing, which is unavoidable in ground based photometry.

We proposed that a careful search for optical microvariability in radio quiet QSOs (RQQSOs)
might reveal something important about their central engines, but our early attempts
led to only marginal detections \cite{gksw93}.  If no rapid variations could be detected down to very small levels, then the idea that small fluctuations in the jets dominate the variable optical emission would be supported, since it was logical to assume that RQQSOs did not possess the jets characteristic of radio loud quasars (RLQs) and blazars;  whereas if microvariability could be detected for RQQSOs, but if it were of a distinctly different character from that of radio loud AGN, then 
fluctuations arising from accretion disks could well be involved in the RQQSOs \cite{man93}.  
To properly explore this question we carefully monitored a group of 27 sources which was
comprised of
samples of RQQSOs, RLQs, CDQs
and BL Lacs with similar distributions in redshift and optical magnitude \cite{gks03}. This
extensive program used improved instrumentation on the 1.0 m  telescope at Nainital, India, and very careful analysis techniques. Over 100 nights of data taken over the course of several years were collected so that most sources were observed for at least 6 hours on a few nights in more than one observing season \cite{sta04a}.

The key result of our campaign was that microvariability could be convincingly detected for all four classes of AGN, including the RQQSOs, when the threshold for detection was about 0.01 mag over the course of a night's observing \cite{gks03} and the differential photometry had errors below that threshold.  Unsurprisingly, blazars were active significantly more often than were RQQSOs.  On the other hand, the following result was somewhat unexpected:
there was no statistically significant difference in the duty cycles
of the RQQSOs, RLQs and CDQs.  For each of these three classes microvariability could be detected in $\sim 20\%$ of the nights the sources were observed, whereas for BL Lacs this duty cycle was $\sim 70\%$.  When this intranight optical variability was broken down by the magnitude of detected variations,
essentially all of the fluctuations $>$ 0.03 mag belonged to BL Lacs, while the smaller variations were
essentially detected equally as often for all four classes \cite{sta04a} \cite{sta04b} \cite{sag04}.  The small duty cycles and always small amplitudes of the microvariability for RQQSOs explained why earlier programs, most of which had worse signal-to-noise and/or sparser sampling and/or shorter nightly monitoring periods and/or less careful data analyses, were unable to clearly find  
RQQSO intranight variability.

The important result that RQQSOs show essentially the same variations as do RLQs indicates that their variations probably have the same origin.  While RQQSO fluctuations could arise from accretion disks, while the stronger variations seen in blazars (here the BL Lacs plus the single high polarization CDQ in our CDQ group) could arise from jets, we argue that a single mechanism could easily account for all of the variations \cite{gks03} \cite{sta04}.  Define the Doppler factor as
\begin{equation}
\delta = [\Gamma (1 - \beta {\rm cos}\theta)]^{-1},
\end{equation}  
with $v = \beta c$, $\Gamma = (1 - \beta^2)^{-1/2}$, and $\theta$ the angle between the jet and our line of sight.  Then the observed flux is
\begin{equation}
S_{\nu,\rm obs} = \delta^{n+\alpha}S_{\nu,\rm em},
\end{equation}	
where $n = 2$ if the emission arises from a continuous jet and $n = 3$ if from a knot 
(usually associated with a shocks in the jet), and where $\alpha$ is the spectral index ($S_{\nu} \propto \nu^{-\alpha}$).  As noted above, the strong rapid variations for blazars seen on longer timescales have long been attributed to such large Doppler boosting from relativistic jets (for $\Gamma \gtrsim$ 10) \cite{bla78}.

We found that such relativistic motions could just as nicely produce the observed
properties of AGN microvariability. The relative fluctuations are most strongly enhanced at the smallest viewing angles. The timescale over which the amplified fluctuation would be observed, $\Delta t_{\rm obs}$, also shrinks as the approaching jet becomes very closely aligned with our point of view, since
$\Delta t_{\rm obs} = \Delta t_{\rm em}/\delta$, thereby making them much easier to detect
in a given night and increasing the observed duty cycle  \cite{gks03}.  In other words, the most probable explanation for
all optical microvariability seen so far is that it arises from relativistically moving plasma in the innermost ($\ll$ 1 pc) portion of the AGN; however, something happens to destroy the jet (or at least eliminate its radio emission) by the $\sim$ 1 pc scale at which VLBI could, but doesn't, detect radio knots in the majority (i.e., the radio quiet class) of AGN. Exactly what mechanism aborts these jets is currently a matter of speculation.

Even though we do not believe that the variations we have detected arise directly from
accretion disks, it is worth noting that they may actually originate in the disks.  This is because disk instabilities can
drive changes in the mass flux entering the jets and/or the velocity of the jets.  Those perturbations would yield the actual changes in $S_{\nu,\rm em}$, which are
in turn amplified by relativistic motions, since the jets are almost certainly launched from the accretion disks \cite{wii06}.
 
\section{CONICAL JETS AND THEIR IMPLICATIONS FOR BLAZARS}

\subsection{Reconciling Fast Jets with Slow Knots}

A growing subset of blazars have been found to emit very high energy (TeV) photons 
\cite{kra01}.
The most probable mechanism for explaining $\gamma$-ray emission involves inverse Compton
scattering of the synchrotron photons off the extremely relativistic electrons in the
jets; this is the synchrotron-self-Compton (SSC) mechanism.  Inverse Compton scattering of 
external photons, arising from the accretion disk or the broad line clouds, may also contribute to the
high energy spectrum.  If TeV energies are reached, very large values (approaching 100) of  the bulk jet Lorentz factor, $\Gamma$,
 are favored by detailed models of
$\gamma$-ray emission \cite{mas97} \cite{kra01}, particularly since the
cross-section for such energetic photons to be absorbed by the intergalactic
IR background to yield $e^+e^-$ pairs is a real constraint.
On the other hand, the radio knots discovered in these TeV blazars have been found to show surprisingly low values of the apparent transverse velocity, with the majority of components apparently subluminal, e.g., \cite{pin04}.  Since the expression for the apparent transverse speed is
\begin{equation}
\beta_{app} = \frac{\beta ~{\rm sin}\theta}{1 - \beta ~{\rm cos}\theta}, 
\end{equation}  
low values of $\beta_{app}$ are not anticipated if $\Gamma$ is very high (so that  $\beta \rightarrow 1$) and $\theta$ is very small $\simeq 1/\Gamma$, as is expected for blazars.
The statistics of these relatively slow component motions would imply that the typical $\Gamma$ factor is only $\sim$2--4, where the usual assumption of a cylindrical jet is made.

This large discrepancy in apparent jet velocities has attracted substantial attention and several
possible explanations for this observational result have been proposed.  One hypothesis is based on the earlier idea that the jet possesses a ``spine--sheath'' structure,
e.g., \cite{chi00}.  If the central core (spine) is ultrarelativistic and thus capable of producing the TeV photons, while the outer layer (sheath) is only mildly relativistic, the sheath could yield the observed radio emission with knots showing only modest apparent velocities  \cite{ghi04}.  A second proposal is that the jet has a strong longitudinal velocity gradient instead of a strong transverse one; i.e., the jet
decelerates dramatically between the sub-pc scale at which the $\gamma$-rays are produced and the $\sim$pc scale at which the radio emission emerges \cite{geo03}.  A third way around this conundrum is to assume that the viewing angle to the source is typically $\theta \ll 1/\Gamma$, but this is statistically very unlikely \cite{pin04}.  While the spine--sheath hypothesis remains viable, the decelerating jet
proposal has energetic problems and also is possibly contradicted by the the result that
blazar jets appear to retain roughly the same Lorentz factor
all the way out to multi-kpc scales when reasonable
estimates can be made, e.g.\ \cite{jor04}.

Recently we \cite{gkdw04} have proposed another alternative, which dispenses with very 
large velocity gradients across the jet, very rapid deceleration, 
or extremely unlikely tiny viewing angles.  Instead we noted that both observations \cite{m87} and theory and simulations indicate that jets, at least on the very small scales of interest,
should have significant opening angles, and are better described as conical than as cylindrical.  Thus we investigated how jets with constant speeds but finite opening angles (full opening angle $\equiv \omega$) would appear to us if the viewing angle, $\theta \simeq \omega$. 

In terms of the emitted flux $S_e$, the observed weighted flux is
\begin{equation}
S_{o,w} = \int_{\Omega} \delta^n(\Omega) ~S_{e}(\Omega)~ d\Omega~ \equiv~ \bar{A}(\theta) S_{e},
\end{equation}
where we have  integrated over the solid angle $\Omega$ corresponding to the opening angle $\omega$ and we have defined the
mean amplification factor, $\bar{A}(\theta)$.
An integration of the (boosted) flux weighted
apparent velocity over the jet cross-section yields the
weighted observed value of the apparent velocity of the jet,
\begin{equation}
{\vec \beta}_{app,w} = 
{\frac{1}{S_{o,w}}} \int_{\Omega} ~{\vec \beta}(\Omega)~
 \delta^n(\Omega)~ dS_{e}(\Omega)~  d\Omega .  
\end{equation}
Note that the resultant vector is along the line joining the
directions of the core and the center of the jet's cross-section.
If $\theta < \omega/2$ then part of the jet is on the ``other side'' of the viewing axis and that part reduces the apparent speed.

The key result that emerges from allowing for the expected finite opening angle of
a jet is that even ultrarelativistic jets would usually appear to be modestly superluminal,
and only very very rarely would a value of $\beta \lesssim \Gamma$ be seen.
For instance, even for the extreme case of
$\Gamma = 100$ and a modest total opening angle,
$\omega = 5^{\circ}$, over 73\%
of the radio components would be detected with $v_{app} < 10c$,
while for $\omega = 10^{\circ}$, over 87$\%$ 
would fall
into this category. Over 41$\%$ (for $\omega = 5^{\circ}$)
and over 69$\%$ (for $\omega = 10^{\circ}$) would 
actually be seen as subluminal sources.  

Thus the predominance of marginally superluminal or
even subluminal motions for VLBI knots
among the TeV blazars does not imply
that their jets cannot be ultrarelativistic.
Instead, a combination of  high  $\Gamma$ factors and 
modest jet opening angles  can just as well
explain the preponderance of low $v_{app}$ values.
At the same time, high $\Gamma$ factor jets ($> 15$)
are demanded to  efficiently produce
the TeV photons by inverse Compton scattering (such relatively
modest values emerge from models only when de-reddening of the TeV spectrum
by the IR background is ignored) and higher values
($\Gamma > 40$) are usually required when the TeV spectrum is 
appropriately de-reddened \cite{kra01}.

\subsection{``Observed'' and actual Lorentz factors}

Very Long Baseline 
Interferometry (VLBI) measurements of the
apparent motions of the parsec-scale radio knots have often been
used to constrain a combination of $\Gamma$ and $\theta$ \cite{ver94}.
The degeneracy can only be broken by combining these data 
with additional observations, such as flux variability, or 
high-energy photons  from  the SSC mechanism.
In these analyses it also has been customary
to assume (often implicitly) a narrow
cylindrical geometry for the jet, but as discussed above, a conical
structure is more reasonable.

We have just extended the conical jet model
to modify the 
procedure that is usually followed to infer $\Gamma$, by combining
the apparent (often superluminal) speed, $v_{app}$, of VLBI
components with the estimated value 
of the bulk Doppler factor, $\delta$, of the jet whose axis makes an 
angle $\theta$ from the line-of-sight \cite{gkwd06}.  
The value of  $\delta$ can be estimated from radio 
observations of flux variability associated with a new VLBI 
component (``knot''), by adopting some maximum physically
attainable value for 
the intrinsic brightness 
temperature, $T_{max}$ \cite{val99}.
This $T_{max}$ is most reasonably set by the equipartition condition 
$\sim 5 \times 10^{10}$ K \cite{rea94}. 
If an appropriate variability timescale, $\tau_{obs}$ is found
corresponding to an observed flux variation $\Delta S$ measured
at a frequency $\nu$, then 
$T_{B,obs} \propto \Delta S/(\tau_{obs} \nu)^2$ and 
$\delta = (T_{B,obs}/T_{max})^{1/(3+\alpha)}$ \cite{ter94}.
This method, which actually produces a lower bound to
$\delta$, or  $\delta_{min}$, has 
been used frequently because it does not require
VLBI measurements. 

Equations (1) and (3) can be combined to solve for $\Gamma$ and 
$\theta$ of the knot in terms of $\beta_{app}$ and $\delta_{min}$;
assuming a cylindrical jet \cite{gue97}
\begin{equation}
\Gamma =  \frac{\beta_{app}^2 + \delta_{min}^2 +1}{2 \delta_{min}}; 
\end{equation}    
\begin{equation}
{\rm tan}~ \theta = \frac{2 \beta_{app}}{\beta_{app}^2 + \delta_{min}^2 -1}. 
\end{equation}  

One can now quantify how the solutions for $\Gamma$ and
$\theta $ (Eqs.\ 6 and 7) are affected when an allowance is 
made for the jet's conical geometry (with a finite full opening angle, 
$\omega$), which, as mentioned 
above, can be several degrees on parsec scales.  
The effective value
(that which would be observed)
of $\delta_e$  is
\begin{equation}
\delta_{e} = \bar{A}^{1/(n+\alpha)}.
\end{equation}
These effective parameters can now be used to compute the 
values of $\Gamma_{\inf}$ and $\theta_{\inf}$ that would be inferred
from the standard formulae and then these inferred values
can then be compared with the actual intrinsic values
of $\Gamma$ and $\theta$ adopted for the jet.

We find that the characteristic behavior of $\Gamma_{\inf}$ depends on
$\theta$.  For $\theta$ less than some critical angle,
$\theta_c \simeq \omega/2$, $\Gamma_{\inf}$ remains essentially
constant at a value which can be much smaller than $\Gamma$.
The computed expectation value is dominated by this reduced
$\Gamma_{\inf}$, since for $\theta > \theta_c$ the probability
of viewing such a source, $p(\theta)$, drops drastically.  
Approaching the critical viewing angle from below,
a sharp rise in $\Gamma_{\inf}$ to a value above $\Gamma$
is found, the amplitude of
which  is more pronounced
for larger $\omega$.  At still larger $\theta > \theta_c$,
$\Gamma_{\inf}$ declines rapidly and asymptotically approaches $\Gamma$;
however, the chance of seeing a source in either of these last
two regimes is very small.
This behavior arises from the spatial sharpness of
the region of strongest Doppler boosting, across which the 
gradients of $\beta_{app}$ and $A$ can be positively or negatively
correlated.

The expectation values, $\langle \theta_{\inf} \rangle$, are 
found to be virtually independent of $\omega$, and we find
 $\langle \theta_{\inf} \rangle \simeq1/1.8 \Gamma$, 
quite a bit smaller than the usually assumed $1/\Gamma$ \cite{gkwd06}
To summarize, the standard procedure of estimating $\Gamma$ and
$\theta$ from flux variability yielding $\delta$ can 
grossly underestimate their values if the jets
are highly relativistic and have modest opening angles.
Often the standard procedure, cf. \cite{jor04} may yield implausibly
precise alignment ($\theta_{\inf} \ll 1^{\circ}$) even
when the true viewing angle (to the axis of the jet)
is a few degrees.

Ultrarelativistic bulk motion in the VLBI jets, 
as argued here to be in accord with a wide variety of observations, 
has other observational implications. 
One such is that the deprojected length of jets as well as the
radio lobe separation could be substantially overestimated, since the actual
viewing angle is often much larger than the value $\theta_{\inf}$, inferred by
assuming the jet to be cylindrical.

\section{THE INTERFACE BETWEEN RADIO GALAXIES AND COSMOLOGY}

There are many reasons for believing that the interaction of AGN with their
environment can have a substantial impact on that environment.  Over the past few
years it has become clear that the energy released by AGN, most likely through the mechanical input from jets,
is sufficient to stop the cooling of gas in many clusters of galaxies which would otherwise produce too much star formation in the giant galaxies at the centers of clusters
of galaxies;  \cite{fab05}
provides a recent review of this topic.
It is also quite clear that the immense energy emitted by AGN over their extensive
lifetimes may have other dramatic impacts on a multi-phase intergalactic medium (IGM).
In particular, we have proposed that the radio lobes inflated by AGN were sufficiently
numerous and large during the ``quasar era'', roughly from redshifts of 1.5 to 3, to
have a significant positive influence on the rate of star formation outside the
central galaxies, the spreading of magnetic fields into the IGM and the injection of metals into the IGM \cite{gkw01} \cite{gkwo02} \cite{gkw03} \cite{gkwb05}.  Several of these conclusions have also been reached independently by
other authors working from different lines of evidence \cite{fur01} \cite{kro01} \cite{sca04} \cite{gne05}.

Our argument for the importance of radio lobe impact is based on a few key points \cite{gkw01}.  Even though powerful RGs and quasars are very rare in the local universe and are not currently capable of making an impact outside their
local clusters, this was not the case if we go back several Gyr.
First, powerful radio galaxies were much more common during the quasar era, with a co-moving density for FR II RGs at those redshifts nearly $10^3$  higher than it is today \cite{wil99}.  The star formation rate in galaxies also seems to have peaked during that epoch, e.g.\ \cite{arc01}, so the possibility that there is a causal connection,
and the direction in which it points, demands an investigation. 

Second, the radio galaxies which are observed in flux limited surveys are only a small fraction of those which are actually active, particularly at higher redshifts.  Because of the combined influence of adiabatic losses,
synchrotron losses and inverse Compton losses against the microwave background photons (which rises in importance as $(1+z)^4$), even intrinsically very powerful jets provide radio powers that can only be detected for a few million years in currently complete catalogs \cite{brw99}.  Hence, many sources may continue to grow to sizes of the order of a megaparsec if the RGs are energized for times exceeding
10$^8$ yr, but they would normally not be detectable at those large sizes.  

Third, for these giant radio lobes to affect a significant fraction of the baryons which will collapse to make new galaxies, they do not need to permeate the entire IGM; rather they need to impinge upon only the fraction of the total volume which is
in the filamentary ``cosmic web'' at $z \sim 2$.  High quality cosmological simulations indicate that
this fraction is just a few percent then \cite{cen99} \cite{cen06}, so that the RGs, which are likely to be triggered in the most massive rapidly forming
galaxies at the intersections of those filaments \cite{gkw01} \cite{gne05}, will be well situated to impact those regions, which we call the ``relevant universe''.  The fourth reason why the fraction of the relevant universe impacted by RGs probably is $\gtrsim$0.1 is that several generations of RGs are expected to be triggered during the $\sim$ 2 Gyr long quasar era.

The hypothesis that radio jets are capable of triggering star formation is an old one \cite{dey81}, and many other people have shown that their impact could be strong on scales
below 100 kpc where it can help explain \cite{ree89} the strong correlation in direction between radio structures and extended emission line regions \cite{mcc89}.   Although the values of gas densities at large distances from active galaxies is not known well enough to allow a firm conclusion, it does appear that
powerful radio jets will remain over-pressured with respect to that external medium out to distances of at least several hundred kpc \cite{gkw01}.  Under these circumstances,
the bow shock at the edge of the cocoon can trigger extensive star formation in an IGM containing cooler, denser, clouds, as shown by several recent
hydrodynamical  and magnetohydrodynamical simulations  \cite{mel02} \cite{fra03} \cite{fra04} \cite{wii04} \cite{cho05} \cite{cho06}.

Since magnetic fields appear to be in rough equipartition with the relativistic plasma throughout the extended radio lobes, where they have typical strengths
of several $\mu$G, the hypothesis that those AGN ejecta provide much of the even weaker magnetic fields which permeate the ICM and cosmic web \cite{ryu98}
is a distinct possibility \cite{gkw01} \cite{kro01} \cite{gkwb05} \cite{kro03}.
   In addition, the discovery of significant abundances of metals in intergalactic clouds, e.g.\ \cite{sca02} demands some way of transporting them
from their source in quite distant galaxies.  Although several other possibilities have been suggested, the idea that these are also carried out with expanding
radio lobes seems to be very attractive  \cite{gkw03} \cite{gkwb05}.

Although the scenario where radio lobes accelerate, and perhaps sometimes trigger, galaxy formation at redshifts $\sim 2$ is very attractive, and
has the other positive attributes discussed in the previous paragraph, it requires a great deal of additional investigation.  The conclusions so far
have been based on using the model for radio source evolution given by Blundell et al.\ \cite{brw99}, which assumes that the typical lifetime
for the central engine activity is 500 Myr.  Since the volume filled by the lobes scales as $\tau_{act}^{18/7}$ \cite{gkw01}, pinning down this
parameter (along with several others) is very important.  

There are other recent models for RG evolution \cite{kda97} \cite{man02}, all of which
purport to provide adequate matches  to the joint distributions radio-power, projected size and redshift seen in low frequency selected catalogs.
Our recent work has shown that none of these models really provide simultaneously good fits to all of these data \cite{bar06}, but that the
better fits are found for 150 Myr $\le \tau_{act}\le$ 250 Myr, implying that the lobes may fill only $\sim 0.1$ of the relevant universe, as opposed to the
$\sim 0.5$ fraction inferred earlier \cite{gkw01}.  We are currently producing modifications to these radio galaxy evolutionary models  which are providing somewhat
better fits to the extensive data sets \cite{barw06}.   Our goal is to extend these models to include the contributions from jetted core sources.
This would allow us to simulate deeper catalogs, such as FIRST and NVSS, that have been produced at higher radio frequencies and where host galaxies and
quasars can be identified through matching to optical catalogs such as the SDSS \cite{ive04}.

\section{CONCLUSIONS}

The immense range of phenomenal phenomena associated with active galaxies has made their study one of the most exciting areas of astrophysics
over the past four decades.  Our understanding of AGN is far from complete, but dramatic strides have been made.  It is now
essentially certain that the emission of radiation and, very often, jets, is associated with accretion onto supermassive black holes.
Studies ranging from the radio through the gamma-ray bands have allowed us to build unification models that work very well (but
not perfectly).  The key conclusion from an immense amount of work  is  that many of the features apparently distinguishing classes
of AGN are due to differences in the mass of the SMBHs, the accretion rates
onto them and the angles at which we happen to view those accretion flows.

Our recent work on microvariability of various classes of AGN certainly supports the orientation-based
unification hypothesis; however, it also implies that relativistic jets emitting optical radiation  are very common on the innermost scales,
but that such jets often do not extend out to the parsec scales where radio knots are seen.
Further extensive monitoring of even larger samples would be very helpful to confirm or challenge this hypothesis;  if confirmed,
a good explanation of how such jets may be quenched is needed.
On a related front, our work has shown that
several difficulties related to blazar jets might be overcome if they are moving faster than usually assumed, and have
bulk Lorentz factors frequently exceeding 30, 
but also have a finite opening angle of a few degrees.  Here too, additional work is needed to see if these assumptions are
consistent will all existing data.

The interplay between active galaxies and the growth of structure in the universe has become a very active topic over
the past few years.  The feedback from AGN onto the gas which is in the process of forming new galaxies may have both
inhibitory and excitatory aspects, and we have focused on the latter.  Supersonically expanding radio lobes cover 
extremely large volumes, and if over-pressured with respect to a multi-phase IGM they can trigger extensive star
formation and perhaps the birth of entire galaxies.  These radio lobes also may have key roles to play in spreading both
magnetic fields and metals throughout the universe, so more detailed theoretical and observational studies of all
of the ramifications of this scenario would be most worthwhile.

\begin{acknowledgments}
I thank my students Paramita Barai and  Eunwoo Choi and my collaborators, particularly Samir Dhurde, Ram Sagar, Chelliah Stalin,
and above all, Gopal-Krishna, for their usually dominant contributions to our joint work described here.
I am most grateful to Myeong-Gu Park and Heon-Young Chang for the kind invitation to present the lectures on which 
this paper is based and for arranging the exceptional hospitality at APCTP in Pohang.  This work is supported in part by a subcontract to
Georgia State University from US NSF grant AST-0507529 to the University of Washington.
\end{acknowledgments}





\begin{thebibliography}{}
\bibitem{geb00} K.\ Gebhardt et al., ApJ {\bf 539}, L13 (2000). 
\bibitem{pet97} B.\ M.\ Peterson, An Introduction to Active Galactic Nuclei (Cambridge, Cambridge Univ. Press) (1997).
\bibitem{kro99} J.\ H.\ Krolik, Active Galactic Nuclei (Princeton, Princeton Univ. Press) (1999).
\bibitem{kem99} A.\ K.\ Kembhavi and J.\ V.\ Narlikar, Quasars and Active Galactic Nuclei (Cambridge, Cambridge Univ.\ Press) (1999).
\bibitem{sey43} C.\ Seyfert, ApJ, {\bf 97}, 28 (1943).
\bibitem{kha74} E.\ Ye.\ Khachikian and D.\ W.\ Weedman, ApJ, {\bf 192}, 581 (1974).
\bibitem{fan74} B.\ L.\ Fanaroff and J.\ M.\ Riley, MNRAS, {\bf 167}, 31P (1974).
\bibitem{bbr84} M.\ C.\ Begelman, R.\ D.\ Blandford and M.\ J.\ Rees, Rev.\ Mod.\ Phys., {\bf 56}, 255 (1984).
\bibitem{aah90} P.\ A.\ Hughes, H.\ D.\ Aller, and M.\ F.\ Aller, ApJ, {\bf 374}, 57 (1991).
\bibitem{wii85} P.\ J.\ Wiita, Phys.\ Rep., {\bf 123}, 117 (1985).
\bibitem{ive02} Z.\ Ivezi{\'c}, et al.\ AJ, {\bf 124}, 2364 (2002).
\bibitem{xxx01} R.\ M.\ Sambruna, L.\ Maraschi and C.\ M.\ Urry, ApJ, {\bf 463}, 444 (1996).
\bibitem{tho74} K.\ Thorne, ApJ, {\bf 191}, 507 (1974).
\bibitem{bla78} R.\ D.\ Blandford and A.\ K{\"o}nigl, ApJ, {\bf 232}, 34 (1979).
\bibitem{bar89} P.\ Barthel, ApJ, {\bf 336}, 606 (1989).
\bibitem{ant83} R.\ Antonucci, Ann.\ Rev.\ Astr.\ Ap., {\bf 31}, 473 (1993).
\bibitem{ant80} R.\ Antonucci and J.\ Miller, ApJ, {\bf 297}, 621 (1985).
\bibitem{urr95} C.\ M.\ Urry and P.\ Padovani, Pub.\ Astr.\ Soc.\ Pacific, {\bf 107}, 803 (1995).
\bibitem{xxx02} S.\ Villeux, D.\ B.\ Sanders and D.-C.\ Kim, ApJ, {\bf 522}, 139 (1999).
\bibitem{xxx03} R.\ G.\ McMahon, R.\ S.\ Priddey, A.\ Omont, I.\ Snellen and S.\ Withington, MNRAS, {\bf 309}, L1 (1999).
\bibitem{hop05} P.\ F.\ Hopkins, L.\ Hernquist, T.\ J.\ Cox, T.\ Di Matteo, B.\ Robertson, and V.\ Springel, ApJ, {\bf 630}, 716  (2005).
\bibitem{gkw00} Gopal-Krishna and P.\ J.\ Wiita, A\&A, {\bf 363}, 507 (2000).
\bibitem{lai94} R.\ A.\ Laing, C.\ R. Jenkins, J.\ V.\ Wall and S.\ W.\ Unger, in G.\ V.\ Bicknell, M.\ A.\ Dopita and P.\ J.\ Quinn, eds., The First Stromlo Symposium: the Physics of Active Galaxies, ASP Conf.\ Ser., Vol.\ 54 (San Francisco, Astr. Soc. Pac.), p.\ 201 (1994).
\bibitem{gkw96} Gopal-Krishna, V.\ K.\ Kulkarni and P.\ J. Wiita, ApJ, {\bf 463}, L1 (1996).
\bibitem{tri06} M.\ Trippe and P.\ J.\ Wiita, in preparation (2006).
\bibitem{har06} M.\ J.\ Hardcastle, D.\ A.\ Evans and J.\ H.\ Croston, MNRAS (submitted),
astro-ph/0603090 (2006).
\bibitem{mil89} H.\ R.\ Miller, M.\ T.\ Carini and B.\ D.\ Goodrich, Nature, {\bf 337}, 627 (1989).
\bibitem{dut96} J.\ A.\ de Deigo, D.\ Dultzin-Hacyan, A.\ Ramirez and E. Benitez, ApJ, {\bf 501}, 69 (1998).
\bibitem{gksw93} Gopal-Krishna, R.\ Sagar and P.\ J.\ Wiita, MNRAS, {\bf 262}, 963 (1993).
\bibitem{man93} A.\ V.\ Mangalam and P.\ J.\ Wiita, ApJ, {\bf 406}, 420 (1993).
\bibitem{gks03} Gopal-Krishna, C.\ S.\ Stalin, R.\ Sagar and P.\ J.\ Wiita, 
ApJ, {\bf 586}, L25 (2003). 
\bibitem{sta04a} C.\ S.\ Stalin, Gopal-Krishna, R.\ Sagar and P.\ J.\ Wiita, J.\ Ap.\ Astr.,
{\bf 25}, 1 (2004).
\bibitem{sta04b} C.\ S.\ Stalin, Gopal-Krishna, R.\ Sagar and P.\ J.\ Wiita, MNRAS,
{\bf 350}, 175 (2004).
\bibitem{sag04} R.\ Sagar, C.\ S.\ Stalin, Gopal-Krishna, and P.\ J.\ Wiita, MNRAS,
{\bf 348}, 176 (2004).
\bibitem{wii06} P.\ J.\ Wiita, in Blazar Variability II: Entering the GLAST Era, J.\ Webb,
H.\ R.\ Miller and K.\ Marshall, eds., ASP Conf. Ser. (San Francisco, Astr.\ Soc.\ Pac.), in press (2006).
\bibitem{kra01} Krawczynski, P.\ S.\ Coppi and F.\ Aharonian, MNRAS, {\bf 336}, 721 (2002).
\bibitem{mas97} A.\ Mastichiadis \& J.\ G.\ Kirk, A\&A, {\bf 320}, 19 (1997).
\bibitem{pin04} B.\ G.\ Piner and P.\ G.\ Edwards, ApJ, {\bf 600}, 115 (2004).
\bibitem{chi00} M.\ Chiaberge, A.\ Celotti, and G.\ Ghisellini, A\&A, {\bf 358}, 104 (2000).
\bibitem{ghi04} G.\ Ghisellini, F.\ Tavecchio and M.\ Chiaberge, A\&A, {\bf 432}, 401 (2005).
\bibitem{geo03} M.\ Georganopoulos and D.\ Kazanas, ApJ, {\bf 594}, L27 (2003).
\bibitem{jor04} S.\ Jorstad \& A.\ P.\ Marscher, ApJ, {\bf 614}, 615 (2004).
\bibitem{gkdw04} Gopal-Krishna, S.\ Dhurde and P.\ J.\ Wiita, ApJ, {\bf 615}, L81 (2004).
\bibitem{m87} W.\ Junor, J.\ A.\ Biretta and M.\ Livio, Nature, {\bf 401}, 891 (1999).
\bibitem{ver94} R.\ C.\ Vermeulen and M.\ H.\ Cohen, ApJ, {\bf 430}, 467 (1994).
\bibitem{gkwd06} Gopal-Krishna,  P.\ J.\ Wiita, and S.\ Dhurde, MNRAS, submitted.
\bibitem{val99} E.\ Valtaoja, A.\ L\"ahteenm\"aki, H.\ Ter\"asranta and M.\
Lainela, 1999, ApJS, {\bf 120}, 95  (1999).
\bibitem{rea94} A.\ C.\ S.\ Readhead, ApJ, {\bf 426}, 51 (1994).
\bibitem{ter94} H.\ Ter\"asranta and E.\  Valtaoja,  A\&A, {\bf 283}, 51 (1994).
\bibitem{gue97} E.\ J.\ Guerra and R.\ A.\  Daly, 1997, ApJ, {\bf 491}, 483 (1997).
\bibitem{fab05} J.\ R.\ Peterson and A.\ C.\ Fabian, Phys.\ Rep., in press, astro-ph/0512549 (2005).
\bibitem{gkw01} Gopal-Krishna and P.\ J.\ Wiita, ApJ, {\bf 560}, L115 (2001).
\bibitem{gkwo02} Gopal-Krishna, P.\ J.\ Wiita,and M.\ A.\ Osterman, in Active Galactic Nuclei: from Central Engine to Host Galaxy, ASP, Conf. Ser. Vol.\ 290, eds.\ S.\ Collin, F.\ Combes and I.\ Shlosman. ASP (San Francisco, Astronomical Society of the Pacific),  p.\ 319 (2003).
\bibitem{gkw03} Gopal-Krishna and P.\ J.\ Wiita in Radio Astronomy at the Fringe, ASP Conference Proceedings, Vol. 300, eds.\ J.\ A.\ Zensus, M.\ H.\ Cohen and E.\ Ros, (San Francisco, Astronomical Society of the Pacific), p.\ 293 (2003).
\bibitem{gkwb05} Gopal-Krishna, P.\ J.\ Wiita, and P.\ Barai, J.\ Korean Astron.\ Soc.\ (2004).  
\bibitem{fur01} S.\ Furlanetto and A.\ Loeb, ApJ, {\bf 556}, 619 (2001).
\bibitem{kro01} P.\ P.\ Kronberg, Q.\ W.\ Dufton, H.\ Li and S.\ A.\ Colgate, S. A., ApJ, {\bf 560}, 178 (2001). 
\bibitem{sca04} E.\ Scannapieco and S.\ P.\ Oh, ApJ, {\bf 608}, 62 (2004). 
\bibitem{gne05} R.\ Levine and N.\ Y.\ Gnedin, ApJ, {\bf 632}, 727 (2005).
\bibitem{wil99} C.\ J.\ Willott, S.\ Rawlings, K.\ M.\ Blundell and M.\ Lacy and S.\ A.\ Eales, 2001, MNRAS, {\bf 322}, 536 (2001).
\bibitem{arc01} L.\ Cowie, A.\ Songaila and A.\ J.\  Barger, AJ {\bf 118}, 603 (1999).
\bibitem{brw99} K.\ M.\ Blundell, S.\ Rawlings and C.\ J.\ Willott, AJ, {\bf 117}, 677 (1999).
\bibitem{cen99} R.\ Cen and J.\ P.\ Ostriker, ApJ, {\bf 570}, 457 (1999).
\bibitem{cen06} R.\ Cen and J.\ P.\ Ostriker, ApJ, in press (astro-ph/0601008) (2006).
\bibitem{dey81} D.\ De Young, Nature, {\bf 293},  43 (1981).
\bibitem{ree89} M.\ J.\ Rees, MNRAS, {\bf 239}, 1P (1989).
\bibitem{mcc89} P.\ J.\ McCarthy,  W.\ J.\ M.\ van Breugel, H.\ Spinrad and S.\ Djorgovski, ApJ, {\bf 321}, L29 (1987).
\bibitem{mel02} G.\ Mellema, J.\ D.\ Kurk and H.\ J.\ A.\ R{\"o}ttgering, A\&A, {\bf 395}, L13 (2002). 
\bibitem{fra03} P.\ C.\ Fragile, S.\ D.\ Murray, P.\ Annios and W.\ van Breugel, ApJ, {\bf 604}, 74 (2004).
\bibitem{fra04} P.\ C.\ Fragile, P.\ Annios, K.\ Gustafson and S.\ D.\ Murray, ApJ, {\bf 619}, 327 (2005).
\bibitem{wii04} P.\ J.\ Wiita, Ap. \& Spa.\ Sci., {\bf 293}, 255 (2004).
\bibitem{cho05} E.\ Choi and D.\ Ryu, New Astr., {\bf 11}, 116 (2005).
\bibitem{cho06} E.\ Choi, P.| J.\ Wiita and D.\ Ryu, in preparation (2006).
\bibitem{ryu98} D.\ Ryu, H.\ Kang, and P.\ L.\ Biermann, A\&A, {\bf 335}, (1998).
\bibitem{kro03} P.\ P.\ Kronberg, J.\ Korean Astron.\ Soc., {\bf 37}, 501 (2004).
\bibitem{sca02} E.\ Scannapieco, A.\ Ferrara and P.\ Madau, ApJ, {\bf 574}, 590 (2002).
\bibitem{kda97} C.\ R.\ Kaiser, J.\ Dennett-Thorpe and P.\ Alexander, MNRAS, {\bf 292}, 723 (1997)
\bibitem{man02} E.\ Manolakou and J.\ G.\ Kirk, A\&A, {\bf 391}, 127 (2002). 
\bibitem{bar06} P.\ Barai and P.\ J.\ Wiita, MNRAS, submitted (2006).
\bibitem{barw06} P.\ Barai and P.\ J.\ Wiita, in preparation (2006).
\bibitem{ive04} Z.\ Ivezi{\'c} et al. in Multiwavelength  AGN Surveys, eds.\ R. M{\'u}jica and R.\ Maiolino
(Singapore, World Scientific)  p.\ 53 (2004).






\end{thebibliography}
\end{document}